\documentclass{PoS}
\usepackage{graphicx}

\newcommand{\kms}{\, \rm km\,s^{-1}} 
\newcommand{\msun}{\,M_{\odot}}

\newcommand{\ergs}{\,{\rm erg\,s}^{-1}}

\newcommand{\mbh}{M_{\rm BH}}
\newcommand{\lbol}{L_{\rm bol}}

\newcommand{\lbedd}{L_{\rm bol}/L_{\rm Edd}}

\title{Are Narrow Line Seyfert 1s a special class of Active Galactic Nuclei?}

\ShortTitle{Are NLSy1s a special class of AGN?}

\author{\speaker{M. Valencia-S.}$^{,1,}$\thanks{Member of the International Max Planck Research School (IMPRS) for Astronomy and Astrophysics at the MPIfR and the Universities of Bonn and Cologne.}\\
        E-mail: \email{mvalencias@ph1.uni-koeln.de}}

\author{J. Zuther$^{1}$, ~A. Eckart$^{1,2}$, ~S. Smajic$^{1,\dagger}$, ~C. Iserlohe$^{1}$, ~M. Garc\'{i}a-Mar\'{i}n$^{1}$, ~G. Busch$^{1,}$\thanks{Member of the Bonn-Cologne Graduate School (BCGS) of Physics and Astronomy.}~,
~M. Vitale$^{1,\dagger}$, ~M. Bremer$^{1}$, ~S. Fischer$^{1}$, ~M. Horrobin$^{1}$, ~L. Moser$^{1}$, Y.E. Rashed$^{1}$, ~and ~C. Straubmeier$^{1}$\\
       \llap{$^{1}$} I. Physikalisches Institut, Universit\"at zu K\"oln, Z\"ulpicher Str. 77, 50937 K\"oln, Germany\\
       \llap{$^{2}$} Max-Planck-Institut f\"ur Radioastronomie, Auf dem H\"ugel 69, 53121 Bonn, Germany}

\abstract{No. Due to their apparently extreme optical to X-ray properties,
         Narrow Line Seyfert 1s (NLSy1s) have been considered 
         a special class of active galactic nuclei (AGN). Here, we summarize 
         observational results from different groups to conclude 
         that none of the characteristics that are typically used to define 
         the NLSy1s as a distinct group -- from the, nowadays called, Broad 
         Line Seyfert 1s (BLSy1s) -- is unique, nor ubiquitous of these particular 
         sources, but shared by the whole Type~1 AGN.
		 Historically, the NLSy1s have been distinguished from the BLSy1s by the narrow width of the
		 broad H$\beta$ emission line. The upper limit on the full width at half maximum
		 of this line is $2000\kms$ for NLSy1s, while in BLSy1s it can be of several
		 thousands of $\kms$. However, this border has been arbitrarily set and does
		 not correspond to the change of any physical property. All observed parameters
		 in Type~1 AGN cover continues ranges of values, which does not allow to
		 infer the existence of two different kind of populations with
		 ${\rm FWHM_{H\beta,broad}}=2000\kms$ as dividing point.
		 We argue that the usage of this velocity limit to define samples of NLSy1s galaxies
		 -- as it is usually done in comparative studies --, together
		 with the well known observational biases, naturally favors the selection 
		 of sources with low black hole masses and high Eddington ratios that
		 are hosted by blue spiral galaxies. 
         Therefore selection biases might be responsible for the 
         reported differences between NLSy1 and BLSy1 sources.}

\FullConference{Nuclei of Seyfert galaxies and QSOs - Central engine \& conditions of star formation,\\
		November 6-8, 2012\\
		Max-Planck-Insitut f\"ur Radioastronomie (MPIfR), Bonn, Germany}

\begin{document}

\section{Introduction}

Irrespective of their luminosity, active galactic nuclei (AGN) have been classified in
two main types according to the possibility of having a clear view
toward the central engine. Those with their symmetry axes close to the line of sight --
indicated by their radio emission and/or the presence of broad lines in the spectrum --
are classified as Type~1, while those at higher inclinations -- suggested by the high
absorption column densities in the X-rays probably originating in dusty clumps %located
at outer regions of the accretion disk -- are considered Type~2 sources.
Unobscured Type~1 sources are ideal targets to study the physics of accretion onto supermassive black holes,
because they allow us to reconstruct the AGN intrinsic spectral energy distributions and
constrain the theoretical models (e.g., \cite{antonucci1,antonucci2} and refs. therein).

In the local universe, Type~1 AGN -- constituted mostly by Seyfert 1s -- span a wide
range of bolometric luminosities ($44 \lesssim \log(\lbol/\ergs) \lesssim 47$),
black hole masses ($6 \lesssim \log(\mbh/\msun) \lesssim 9$)
and Eddington ratios ($-2 \lesssim \log(\lbedd) \lesssim 1$). These objects show
broad recombination emission lines with typical widths of the order of several thousands of $\kms$.  
Osterbrock \& Pogge (1985) identified a group of AGN with ``all the properties of
Seyfert 1 or 1.5 galaxies, but unusually [by that time] narrow H{\sc i} lines'', which they
called Narrow Line Seyfert 1s (NLSy1s) \cite{osterbpogge}. %The formal initial criteria were to have ``narrow permitted lines only slightly broader than the forbidden lines'', and [OIII]/H$\beta<3$. 
Later, the quantitative limit ${\rm FWHM_{H\beta, broad}} \leq 2000 \kms$ was
adopted \cite{osterbpogge87}. Other properties of % \cite{goodrich}
these objects are the relatively strong Fe{\sc ii} emission, soft X-ray excess,
steep 2-10\,keV power-law (with photon index $\Gamma \gtrsim 2$), and
short term X-ray variability. %How is the line-width related with that set of peculiar properties?
%Indeed, as
The extreme NLSy1 properties, as it has been discussed extensively in the literature,  
seem to be driven by the Eddington ratio -- as a surrogate of the unobservable mass-accretion
rate --, the black hole mass (e.g., \cite{sulentic, boroson02,dongfe}), and probably
also by the conditions in the narrow line region (e.g., \cite{xu,popovik}).   

%IRAS~01072+4954 is a NLSy1 candidate. In the X-rays, its apearence is typical of
%low-$\mbh$ Type~1 sources \cite{panessa05}, while in the optical, it does not display
%any signs of broad hydrogen emission. Moreover, according to Moran et al. (1996), like
%other composite sources, IRAS~01072+4954 [OIII]$\lambda\lambda 4959,5007$ lines are
%``significantly broader than all other narrow lines in the spectrum, forbidden or permitted'' \cite{moran}.
%Although, this is in contradiction to the main NLSy1 classification criteria, it is
%possible to show that broad line emission is expected, and therefore the available optical
%spectrum might suffer from stellar contamination or low sensitivity.

In the following, we examine some of the properties ascribed to NLSy1s in
order to enlighten the discussion of whether these sources constitute a particular class of AGN, 
or are just normal members of the Type~1 population. 
Reviewing the related literature in the past couple of decades,
and using published available data of thousands of Type~1 sources,  we show that 
the properties used to characterize and define the NLSy1s as a special group of AGN 
neither are shared by all the members, nor they are exclusive of this kind of sources.  
We discuss some observational biases that affect the selection of samples,
and that might also lead to conclude that NLSy1s are different from other Type~1 AGN in comparative studies.
Throughout the text, we use the term Broad Line Seyfert 1s (BLSy1s) to designate Type~1
sources with broad hydrogen-line widths larger than $2000\kms$. The term NLSy1s
is employed following the designation of the authors in each contribution. In that
sense, it is not uniformely defined and e.g., one finds that a source classified as NLSy1 by
one group is called low-$\mbh$ AGN (with broad line emission) by another. 

%%%%

\section{Properties of Narrow Line Seyfert 1s (only?)}

From statistical studies of Type~1 samples, relations between different line emission parameters have been derived.
NLSy1s seem to lie at one extreme of some of those relations, however a direct interpretation
is not trivial essentially for two reasons:
i) The sample selection. Groups of sources that span narrower parameter ranges display ambiguously
correlated data, and cluster in smaller regions on the diagrams than `control' samples
with wider parameter ranges. 
When selecting a sample, the narrower the limit imposed to the broad line width is,
the smaller the ranges in e.g., $\mbh$, $\lbol$, and consequently $\lbedd$ ~become,
as we discuss later. 
ii) The choice of parameters to derive possible trends. Correlations between parameters
that are intimately related to each other can mimic real physical trends
and mislead the interpretation of the data (\cite{veron,antonucci2}).
To these caveats, one has to add the fact that data is treated in different ways by different groups.
Hence, some of the correlations are not confirmed by different samples.
For these reasons, we leave the discussion on correlations and trends for other ocassion,
and concentrate here only on the most common properties used to describe -- and to select -- NLSy1s.

%%%
{\bf Hydrogen beta broad-line is narrow: ${\rm FWHM} \leq 2000\kms$.} 
The most commonly used selection criterium for NLSy1s is the width of the broad H$\beta$ line.
In the literature, the $2000\kms$ limit appears for the first time in Osterbrock
\& Pogge (1987)\cite{osterbpogge87}\footnote{Thought the most cited references are \cite{osterbpogge,goodrich}!}. 
Although from the beginning this limit was recognized as arbitrary and unphysical \cite{goodrich},
it is still used to derive ``physically motivated'' differences between NLSy1s and BLSy1s.
Sources with broad-line widths larger than $2000\kms$, but that have visible Fe{\sc ii} emission
and/or soft X-ray excess should also be considered as members of the NLSy1 group
\cite{veron,goodrich} (and refs.~therein). To account for that, a luminosity-dependent definition of
NLSy1 has been suggested several times (e.g. \cite{laor,veron,dultzin}).
On the other hand, the broad line profiles are not all the same, one (or two) Gaussians (or Lorentzians)
are required to model them \cite{veron,peterson}. The classification of Type~1s
into population A or B sources -- that has a flexible limit at ${\rm FWHM_{H\beta, broad}} \sim 4000 \kms$
-- seems to account for this difference and for other observed properties \cite{sulentic}.
Broad lines can also be narrowed due to the inclination of a ring-like broad line
region \cite{gaskell}.
However, the effects of the AGN orientation might not be equally important in all NLSy1s, 
as has been pointed out in particular sources \cite{peterson,collin}, and also using 
larger samples \cite{boroson}.

%%%
{\bf Emission-line ratios as indicators of the ionizing mechanism: [O{\sc iii}]$\lambda 5007$/H$\beta<3$.} 
The denominator of this fraction refers to the total H$\beta$ emission.
In cases where the broad line region is obscured, like in Seyfert~2 galaxies, this ratio
increases up to $\sim 5-10$. However, for Seyfert~1s that do not suffer extinction\footnote{
Even in cases where the broad line region is obscured by the dust in the host galaxy
-- not that from the torus --, [O{\sc iii}]$\lambda 5007$/H$\beta<3$, like in NGC 7172 \cite{semir}.}
it remains $\lesssim 3$ (e.g., \cite{veron}), and therefore it seems to be a condition of
all unobscured Type~1 AGN\footnote{We verify this fact using the sample of Seyfert 1s
and QSOs of Dong et al. (2011). Out of 4178 objects only one, SDSS~J015142.72+132003.3,
has [O{\sc iii}]/H$\beta>3$ ($=4.29$) \cite{dongfe}. Note that the sample selection in this case favored sources
with minimal galaxy contamination. In other samples, where fitting the stellar component
in the optical is required, this criterium is full filled by $\gtrsim 90$\% of the sources, 
as we corroborate using the measurements of Zhou et al. (2006) and Dong et al. (2012) \cite{zhou,dong2012}.}.
When comparing the [O{\sc iii}] emission with the narrow H$\beta$ component, NLSy1s span
similar ranges as BLSy1s, but some objects also have [O{\sc iii}]/H$\beta$
and [N{\sc ii}]/H$\alpha$ narrow-line ratios in the ranges of composite and
star-forming galaxies (according to the classification by Kewley et al. 2001).
Although it has been argued that this intrusion in the regions occupied by H{\sc ii} galaxies 
in the optical diagnostic diagrams might be an effect of the fitting procedure 
used to model the recombination lines\footnote{V\'eron-Cetty et al. (2001)
show that fitting the broad components with Lorentzians and the narrow ones with
Gaussians in their sample of $\sim 60$ NLSy1s produce an overall shift in the
 distribution of these sources in the [O{\sc iii}]/H$\beta$ vs. [N{\sc ii}]/H$\alpha$
diagram. In this way, only $\sim 10$\% instead of $\sim 30$\% of their sources are
out of the Seyfert region.},
it has been shown by independent studies that this can also be a combined effect of
the low-$\mbh$ and dilution by the stellar component. For example, in the
intermediate-$\mbh$ sample of Xiao et al. (2011), 30\% of the objects lie under
the Seyfert/star-formation division \cite{xiao}.
The authors interpret this as a dilution effect because the median redshift
of the H{\sc ii} sources is somewhat larger ($z\sim 0.1$) than that of the whole sample ($z\sim0.08$).
Interestingly, Stern \& Laor (2012) find that the fraction of Type~1 AGN classified as
star forming sources (in the optical diagrams) increases when decreasing $\mbh$:
from 6\% at $\log (\mbh/\msun) = 8.8$ to 32\% at  $\log (\mbh/\msun) = 6.3$ (Fig.~B1 in \cite{sternlaor}).

%\smallskip

{\bf Strong Fe{\sc ii} optical emission}. Fe{\sc ii} multiplets are present in the spectra of Type~1 AGN 
from the UV to the infrared \cite{borosong,ardila}. 
The detection of optical Fe{\sc ii} was the first property that allowed the identification
of NLSy1 candidates among the Seyfert~2s in the available surveys\cite{osterbpogge87}.
The strength of the iron emission is usually quantified %in the optical
as the ratio $R_{4570}$ between the fluxes (or equivalent widths) of the Fe{\sc ii} blend centered
at $\lambda4570$ and the H$\beta$ broad component. NLSy1s were initially found to have
$R_{4570} \gtrsim 0.5$. However, some research groups have pointed out that 
NLSy1s might not be strong Fe{\sc ii} emitters, but faint H$\beta$ sources (\cite{veron} and refs.~therein).
Using the measurements of Dong et al. (2011)\footnote{In order to allow comparison 
with other samples, we use the total iron flux, i.e., we sum up the contributions
 of narrow and broad Fe{\sc ii} fitted by the authors \cite{dongfe}.}
in a sample of $\sim 4000$ Type~1 AGN, we find that both statements are probably correct (see Fig.~1).
We selected random subsamples of $\sim 200$ objects in three ranges of ${\rm FWHM_{H\beta,broad}}$:
$1500-2000$, $5000-6000$, and $>8500\kms$.

\begin{figure}[t]
    \begin{minipage}[h]{210pt}%{275pt}
       \includegraphics[width=210pt]{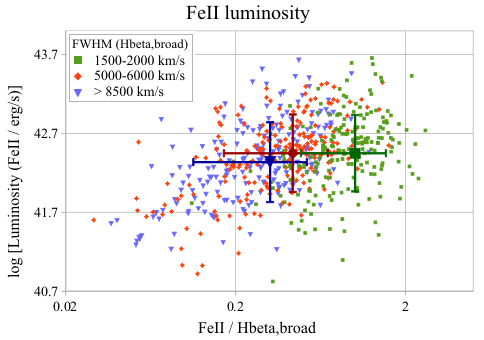}%, height=154
    \end{minipage}
      \hspace*{10pt}
    \begin{minipage}[h]{210pt}
    	\includegraphics[width=210pt]{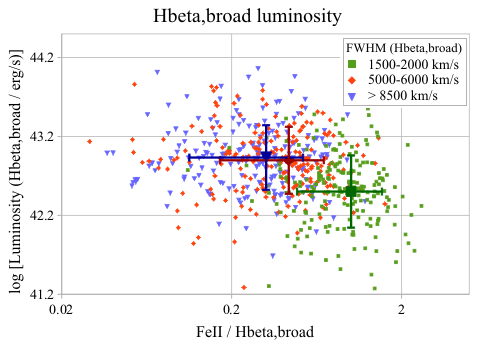}
    \end{minipage}
\caption{\small
Iron and H$\beta$,broad luminosities in Type 1 AGN. The symbols(colours) represent
sources having ${\rm FWHM_{H\beta,broad}}$ within certain ranges, as shown in the
caption. Here, only 200 (randomly selected from the sample of Dong et al. 2011)
objects from each group are shown. Big symbols and bars represent the mean and standar
deviation of the distribution of luminosities and $R_{4570}$ for each group.   
}
\label{fig:firsb}
\end{figure}

We found that, on average, the Fe{\sc ii} luminosity stays approximately constant 
over the first two intervals $\langle\log(L_{\rm Fe{\textsc {ii}}}/\ergs)\rangle\sim 42.45$ and drops
in the last one ($\sim 42.35$), while the luminosity of the H$\beta$ broad
component is lower in the first group of sources $\langle\log(L_{\rm H\beta,broad}/\ergs)\rangle\sim 42.50$
than in the other two ($\sim 42.90$)\footnote{It must be noticed that the measurement uncertainties
of the Fe{\sc ii} flux are $\sim  15$\% \cite{dongfe}, which corresponds to $\sim 0.08\,{\rm dex}$.
However, if the difference in $\langle L_{\rm Fe{\textsc {ii}}} \rangle$ between the three intervals
were dominated by this kind of uncertainties, one would expect the average luminosity values
to scatter around a mean value when one uses different subsamples of randomly selected 200 objects
 in each interval to estimate the average.
In this case instead, we see a systematic shift in $\langle L_{\rm Fe{\textsc {ii}}} \rangle$
from the first two ${\rm FWHM_{H\beta,broad}}$-bins to the last one.  
The measurement uncertainties in the H$\beta$ broad-line fluxes are only $\sim 8$\% \cite{dongfe}.}.
On the other hand, the large range of $R_{4570}$ seen in sources with low-${\rm FWHM_{H\beta,broad}}$ 
is due to the higher spread in the luminosity measurements of the H$\beta$ broad component ($\sim0.5$\,dex) 
compared to that in other Type~1s ($\sim0.2$\,dex) -- is this broad $R_{4570}$-range
related to inclination effects?  
As a result, not all low-${\rm FWHM_{H\beta,broad}}$ are high-$R_{4570}$ sources.
In the NLSy1 sample of Zhou el at. (2006), the authors
point out the presence of objects with very faint or non-detectable Fe{\sc ii} emission,
which they called Fe{\sc ii}-deficient NLSy1s \cite{zhou}. 

{\bf Steep soft X-ray spectrum (soft X-ray excess) and short-term X-ray variability}.  
In the soft X-rays (i.e. below $\sim 1$\,keV), NLSy1s seem to be brighter and more variable
than BLSy1s (e.g. \cite{sulentic,veron,laor} and ref.~therein). This has been associated
with high Eddington ratios. However, not all NLSy1s behave the same.
For example, Panessa et al. (2011) found that ``hard [$>20$\,keV] X-ray selected NLSy1s
do not display particularly strong soft excess emission''\footnote{The authors also pointed out
that ``...indeed only one source [out of 14], IGRJ $19378-0617$, shows a dominant and
strongly variable soft-X-ray component''.} \cite{panessanlsy1}. Similar results had been
reported by Williams et al. (2002, 2004). They noticed that a substantial number of
optically bright, low-redshift NLSy1s did not have soft X-ray counterparts detected by ROSAT.
Follow-up Chandra observations of 17 sources (with NLSy1 characteristics, i.e. 
${\rm FWHM_{H\beta,broad}}<2000\kms$ and $R_{4570}>0.5$) 
showed that their soft X-ray indices ($\Gamma=1.1-3.4$) ``extended far
below those normally observed in NLSy1s''\cite{williams04}. On the other hand,
the extreme variability on short-time scales observed in the soft X-rays seems to be
related to the comparatively low black hole masses found in these AGN rather than to the
possibly high Eddington ratios \cite{dewangan,ai11}.
Ai et al. (2011) suggested the existence of two kinds of low-$\mbh$ AGN: those NLSy1-like
with strong Fe{\sc ii}, soft X-ray excess and high-$\lbedd$, and others more similar to BLSy1s,
i.e., weak Fe{\sc ii} and non-ubiquitous soft X-ray excess. 
%\footnote{However, notice that Type~1 AGN with broad H$\beta$ lines of $\sim5000\kms$ are not Fe{\sc ii} weak, but rather H$\beta$ bright emitters as we showed before.}.
However, in the NLSy1 sample of Zhou et al. (2006) about 15\% of the objects 
have weak iron strength, but high Eddington ratio (i.e. $R_{4570}<0.5$ and $\lbedd>0.5$).
Another case is the source IRAS 01072+4954 with no detected optical Fe{\sc ii}, but $\lbedd>0.2$ \cite{me}.

\section{Do comparative studies suffer from selection biases?}

Here, we briefly describe how a sample selection that uses the width of the broad hydrogen 
lines as a criterium produces a bias in the physical properties of the selected AGN 
that affects the results of comparative studies between NLSy1s and BLSy1s.

Accepting that the broad line region is virialized \cite{peterson} and that its radius 
depends on the optical continuum luminosity at 5100\AA ~(after starlight correction) as
$r_{\rm BLR} \propto L_{5100}^{\sim0.5}$ \cite{kaspi}, it is easy to find that
$(\mbh/\msun) \approx \lbedd\, (f/592.5)^2 \, ({\rm FWHM_{H\beta,broad}}/\kms)^4 $,
where $f$ is the scaling factor. Therefore, selecting sources with ${\rm FWHM_{H\beta, broad}} <2000 \kms$
imposes a maximum limit on the black hole masses of $\sim 2.5\times 10^7\msun$ when $\lbedd=1$ and the
typical value $f=0.75$ are assumed\footnote{Correcting by the broad line profile 
(Gaussian or Lorentzian) can result in one order of magnitude higher black hole 
masses. See \cite{collin} for details.}.
An upper boundary on $\mbh$ also implies an upper limit for the bolometric luminosity
of $\sim (3-10) \times 10^{45}\ergs$ (even when allowing super-Eddington sources).
At these luminosities, low and high Eddington-ratio AGN are detected in complete samples, 
i.e. depending on their $\mbh$, both BLSy1s and NLSy1s are observed. 
However at $\mbh {\rm \sim few} ~10^6\msun$, or correspondingly $\lbol \sim 10^{44}\ergs$,
observational biases only favor the detection of high Eddington sources.
While this kind of selection bias is extensively discussed by Hopkins et al. (2009) \cite{hopkins},
there are indications that the dilution effects on low-$\mbh$ sources might be even more
severe particularly at optical/infrared bands -- i.e. where most of the AGN
with narrow broad components are identified -- than it is estimated there.
If the $\mbh$ vs. bulge-mass relation is not a single power law across the whole
black hole mass range, but is steeper at low black hole masses -- as it has been 
suggested and observed in samples of nearby objects \cite{laormass,graham} --,
then dilution effects become $10-100$ times more severe at $10^6-10^5 \msun$
than previously estimated\footnote{The dilution of the AGN signatures by the stellar
emission of the host galaxy can be estimated as
the contrast between the AGN ($L_{\rm AGN,B}\sim \lbol/9$) and the host luminosities, in the B-band
$L_{\rm AGN,B}/L_{\rm host,B} = \frac{1}{3} \times 10^3\, (\lambda_{\rm Edd}/0.1)\,(\mbh/\msun)\,(L_{\rm host,B}/L_{\rm \odot,B})$, with $\lambda_{\rm Edd} \equiv \lbedd$. The relation between the
black hole and the bulge mass for a source with $\mbh\sim10^7\msun$ is $\mbh \sim 0.001 M_{\rm bulge}$. 
However for lower-$\mbh$ Sersic galaxies, following the relation found by Scott et al. (2013),
at $\mbh =10^6\msun$, $\mbh \sim 10^{-4} M_{\rm bulge}$ \cite{graham}.}.  
Hopkins et al. also show that samples with an implicit high Eddington-ratio selection
--like those that are produced when selecting only narrow H$\beta$ broad-line emitters --
favor low-mass disk-dominated host galaxies, as has been found in the case of NLSy1s. 

As an example, we compiled information about the initial group of
NLSy1 sources studied by Osterbrock \& Pogge (1985) \cite{osterbpogge}. We noticed that
all of them have relatively low black hole masses $\log(\mbh/\msun) \sim 5.8-7.2$,
and that most of them (9/11)\footnote{The other two sources: Mrk~1388 seems to be a
high-ionization Seyfert~2 galaxy \cite{osterbrock1388,gordon},
and Mrk~684 presents broad H$\beta$ and strong Fe{\sc ii} emission, but its 
Eddington ratio $\lbedd \sim 10^{-7}$ is very unusual for a Type~1 source 
($\mbh$ and $\lbol$ from \cite{wanglu, grupe}, see spectrum in \cite{vaughan}).}
also have high Eddington ratios $\lbedd > 0.4$. As it was mentioned before,
also the calibration and fitting procedures, even when using data from the same survey,
can introduce biases in the sample selection. 
For example, the very careful selection process of low-$\mbh$ sources applied by
Dong et al. (2012) allowed them to detect AGNs with lower Eddington ratios
($\langle \lbedd\sim \rangle -0.7$ versus $-0.4$) 
than previously published samples with similar selection criteria \cite{dong2012}.

%In the case of IRAS~01072+4954, when comparing reddening corrected narrow emission lines [OIII]$\lambda 5007$/H$\beta\sim 1$, which places it in the region populated by composite galaxies. When the expected broad H$\beta$ component (derived as shown in Sect.~5.1 of \cite{me1}) is added up to the corrected narrow H$\beta$, then the ratio of [OIII] to total H$\beta$ fluxes is $\sim 0.2-0.5$ indicating a Sy~1.2 classification according to the definition of Winkel (1992).

\section{Final comments}

Narrow Line Seyfert 1 have been considered a special class of AGN because of their
apparently extreme properties.
After an extensive literature review (not covered in the References of this short
paper due to space  restrictions) and studying some statistical properties of recently
published samples of Type~1, NLSy1s, and low-$\mbh$ sources \cite{dongfe,zhou,dong2012},
we find that apart from the width of the broad H$\beta$ component, which is an arbitrarily defined
limit, none of the other properties is unique, nor ubiquitous of the NLSy1s. Moreover, these sources
seem to be normal Type~1 AGN that naturally cover the lower end of the ${\rm FWHM_{H\beta,broad}}$ range.

Implicit selection effects (${\rm FWHM_{H\beta, broad}} < 2000 \kms$) in NLSy1s samples
can account -- at least partially -- for the following averaged observed properties 
when they are analyzed in comparison to those of BLSy1s: lower $\mbh$,
faster soft X-ray variability, higher $\lbedd$, lower $\lbol$, higher bolometric corrections,
lower redshifts, bluer disk-dominated hosts, $\mbh$ below the bulge mass (or bulge luminosity)
correlations and thereby higher importance of secular evolution processes.
 
On the other hand, it is possible to define a `special' subgroup of sources that meet a certain
set of observational requirements (in ${\rm FWHM_{H, broad}}$, $R_{4570}$, $\Gamma$, $\lbol$-correction,
radio-loudness, narrow line ratios, host galaxy morphology...), but that might be of
little physical interest. Selecting samples based on physical properties would be probably
more helpful to understand the accretion phenomena. In the words of Goodrich
(1989)\footnote{This is Goodrich's reading of the review made by Lawrence (1987) on the
AGN classification and its meaning.} "...a taxonomic system, based not necessarily
on the most easily observed parameters but rather on the most physically relevant
quantities and some theoretical picture, is required in order to make substantial
progress in our interpretation of AGN" \cite{goodrich}.

%Are NLSy1s active nuclei with low mass black holes and high Eddington ratio?
%
%If NLSy1s are defined solely by the width of their broad recombination components
%(which is totally arbitrary), then it is clear that creating narrow broad-components
%in an average luminosity AGN requires low-$\mbh$.  The high $\ledd$ ratio might is a
%collection of effects: i) An observational bias, because at a given $\mbh$ brighter AGN are more probable to
%be detected in magnitude-limited samples than fainter ones. ii) The cosmic downsizing
%of BH activity, according to which at lower redshifts lower-$\mbh$ are more active.
%iii) Dilusion of AGN signatures
%by stellar light, which affects stronger low-$\mbh$ (than high-$\mbh$) sources  with low-$\ledd$
%-- as discussed in more detail further down. An example of this effect can be seen
%in the $\ledd$ ranges covered by the low-$\mbh$ samples of Greene \& Ho (2007) and
%Dong et al (2012). Both of them are selected from the SDSS DR4 and cover similar ranges
%in $\mbh$, but because in the later fainter H$\alpha$ emitters were detected, its
%$\ledd$ distribution is broader and has a lower median (see Fig.~4 in \cite{dong2012}).
%
%
%
%
%Over classification of sources?
%
%Validity of the BPT diagrams on lowlum lowmbh lowioniz sources?
%
%Do NLSy1s prefer pseudobulge galaxies?
%
%
%Selection bias (prob relat with high lambdaedd? at low mbhs)
%
%Separate the source of the qsar like emission from the host Fe{\sc ii}?
%

\end{document}